\documentclass[twocolumn]{aastex63}
\usepackage{natbib}
\usepackage{amsmath} 
\newcommand{\angstrom}{\text{\normalfont\AA }{ }}
\newcommand{\angstromcomma}{\text{\normalfont\AA }{}}
\usepackage[encapsulated]{CJK}

\usepackage{rotating}

\usepackage{float}
\usepackage{textcomp}
\usepackage{makecell}

\begin{document}

\date{\today}


\shortauthors{Setton et al.}

\title{Little Red Dots at an Inflection Point: Ubiquitous ``V-Shaped" Turnover Consistently Occurs at the Balmer Limit}

\shorttitle{Little Red Dot Slopes Inflect at $H\infty$}

\author[0000-0003-4075-7393]{David J. Setton}\thanks{Email: davidsetton@princeton.edu}\thanks{Brinson Prize Fellow}
\affiliation{Department of Astrophysical Sciences, Princeton University, 4 Ivy Lane, Princeton, NJ 08544, USA}

\author[0000-0002-5612-3427]{Jenny E. Greene}
\affiliation{Department of Astrophysical Sciences, Princeton University, 4 Ivy Lane, Princeton, NJ 08544, USA}

\author[0000-0002-2380-9801]{Anna de Graaff}
\affiliation{Max-Planck-Institut f\"ur Astronomie, K\"onigstuhl 17, D-69117, Heidelberg, Germany}

\author[0000-0002-0463-9528]{Yilun Ma  (\begin{CJK*}{UTF8}{gbsn}马逸伦\ignorespacesafterend\end{CJK*})}
\affiliation{Department of Astrophysical Sciences, Princeton University, 4 Ivy Lane, Princeton, NJ 08544, USA}

\author[0000-0001-6755-1315]{Joel Leja}
\affiliation{Department of Astronomy \& Astrophysics, The Pennsylvania State University, University Park, PA 16802, USA}
\affiliation{Institute for Computational \& Data Sciences, The Pennsylvania State University, University Park, PA 16802, USA}
\affiliation{Institute for Gravitation and the Cosmos, The Pennsylvania State University, University Park, PA 16802, USA}

\author[0000-0003-2871-127X]{Jorryt Matthee}
\affiliation{Institute of Science and Technology Austria (ISTA), Am Campus 1, Klosterneuburg, Austria}

\author[0000-0001-5063-8254]{Rachel Bezanson}
\affiliation{Department of Physics and Astronomy and PITT PACC, University of Pittsburgh, Pittsburgh, PA 15260, USA}

\author[0000-0002-3952-8588]{Leindert A. Boogaard} 
\affiliation{Leiden Observatory, Leiden University, PO Box 9513, NL-2300 RA Leiden, The Netherlands}

\author[0000-0001-7151-009X]{Nikko J. Cleri}
\affiliation{Department of Astronomy \& Astrophysics, The Pennsylvania State University, University Park, PA 16802, USA}
\affiliation{Institute for Computational \& Data Sciences, The Pennsylvania State University, University Park, PA 16802, USA}
\affiliation{Institute for Gravitation and the Cosmos, The Pennsylvania State University, University Park, PA 16802, USA}

\author[0000-0003-1561-3814]{Harley Katz}
\affiliation{Department of Astronomy \& Astrophysics, University of Chicago, 5640 S Ellis Avenue, Chicago, IL 60637, USA}
\affiliation{Kavli Institute for Cosmological Physics, University of Chicago, Chicago, IL 60637, USA}

\author[0000-0002-2057-5376]{Ivo Labbe}
\affiliation{Centre for Astrophysics and Supercomputing, Swinburne University of Technology, Melbourne, VIC 3122, Australia}

\author[0000-0003-0695-4414]{Michael V.\ Maseda}
\affiliation{Department of Astronomy, University of Wisconsin-Madison, 475 N. Charter St., Madison, WI 53706 USA}

\author[0000-0002-2446-8770]{Ian McConachie}
\affiliation{Department of Astronomy, University of Wisconsin-Madison, 475 N. Charter St., Madison, WI 53706 USA}

\author[0000-0001-8367-6265]{Tim B. Miller}
\affiliation{Center for Interdisciplinary Exploration and Research in Astrophysics (CIERA), Northwestern University, 1800 Sherman Ave, Evanston, IL 60201, USA}

\author[0000-0002-0108-4176]{Sedona H. Price}
\affiliation{Department of Physics and Astronomy and PITT PACC, University of Pittsburgh, Pittsburgh, PA 15260, USA}

\author[0000-0002-1714-1905]{Katherine A. Suess}
\affiliation{Department for Astrophysical \& Planetary Science, University of Colorado, Boulder, CO 80309, USA}

\author[0000-0002-8282-9888]{Pieter van Dokkum}
\affiliation{Astronomy Department, Yale University, 219 Prospect St,
New Haven, CT 06511, USA}

\author[0000-0001-9269-5046]{Bingjie Wang (\begin{CJK*}{UTF8}{gbsn}王冰洁\ignorespacesafterend\end{CJK*})}
\affiliation{Department of Astronomy \& Astrophysics, The Pennsylvania State University, University Park, PA 16802, USA}
\affiliation{Institute for Computational \& Data Sciences, The Pennsylvania State University, University Park, PA 16802, USA}
\affiliation{Institute for Gravitation and the Cosmos, The Pennsylvania State University, University Park, PA 16802, USA}

\author[0000-0001-8928-4465]{Andrea Weibel}
\affiliation{Department of Astronomy, University of Geneva, Chemin Pegasi 51, 1290 Versoix, Switzerland}

\author[0000-0001-7160-3632]{Katherine E. Whitaker}
\affiliation{Department of Astronomy, University of Massachusetts, Amherst, MA 01003, USA}
\affiliation{Cosmic Dawn Center (DAWN), Denmark}

\author[0000-0003-2919-7495]{Christina C.\ Williams}
\affiliation{NSF National Optical-Infrared Astronomy Research Laboratory, 950 North Cherry Avenue, Tucson, AZ 85719, USA}

\submitjournal{ApJ}

\begin{abstract}

Among the most puzzling early discoveries of JWST are ``Little Red Dots"  -- compact red sources that host broad Balmer emission lines and, in many cases, exhibit a ``V shaped" change in slope in the rest-optical. The physical properties of Little Red Dots currently have order-of-magnitude uncertainties, because models to explain the continuum of these sources differ immensely. Here, we leverage the complete selection of red sources in the RUBIES program, supplemented with public PRISM spectra, to study the origin of this ``V shape". By fitting a broken power law with a flexible inflection point, we find that a large fraction (20/44, nearly all spatially unresolved) of extremely red H$\alpha$ emitters at $2<z<6$ exhibit a strong change in slope, and that all strong inflections appear associated with the Balmer limit ($0.3645$ $\mu$m). Using a simple model of a reddened AGN with an unobscured scattered light component, we demonstrate that the observed ``V shape" in Little Red Dots is unlikely to occur at any specific wavelength if the entire continuum is dominated by light from a power law AGN continuum. In contrast, models with an intrinsic feature at the Balmer limit, such as those that are dominated by evolved stellar populations in the rest-UV-to-optical, can produce the observed spectral shapes, provided that a reddened component picks up sufficiently redward of the break. While no model can comfortably explain the full Little Red Dot spectral energy distribution, the common inflection location suggests that it is most likely a single component that consistently dominates the rest-UV-to-optical in Little Red Dots, and that this component is associated with $T\sim10^4$ K hydrogen due to the clear preference for a break at H$_\infty$.

\end{abstract}



\section{Introduction} \label{sec:intro}


One of the most intriguing discoveries in the first few years of JWST observations has been large numbers of compact red sources \citep[e.g.,][]{Barro2024,Labbe2023UB,Labbe2023LRD}. Follow-up spectroscopy reveals that these sources often contain broad Balmer emission lines with FWHM$>2000$~km\,s$^{-1}$ \citep{Larson2023, Ubler2024, Greene2024, Killi2023, Wang2024_BRD, Wang2024_UB, Kokorev2023_LRD}. At the same time, spectroscopically identified broad-line objects are commonly red and compact \citep{Kocevski2024,Harikane2023,Lin2024}, leading \citet{Matthee2024} to dub this new class of sources ``Little Red Dots". Not only do the sources share a compact morphology (being effectively unresolved in the rest-frame optical, implying $r_e \lesssim 100$ pc), many sources have a characteristic ``V-shape'', wherein the red rest-frame optical slope transitions to a blue rest-frame UV slope. These objects have become the subject of intense study because they are so common; they account for a few percent of $>L_*$ galaxies, and at least 20\% of active galactic nuclei (AGN) with broad lines \citep{Harikane2023}. 

Early photometric papers favored the hypothesis that these objects are AGN based on their compact morphology \citep[e.g.,][]{Onoue2023,Furtak2024,Labbe2023LRD}, later supported by early spectroscopic papers that used the presence of broad emission lines as support for an AGN hypothesis \citep{Furtak2024,Ubler2024,Matthee2024}. However, there has always been some ambiguity about whether the red continuum is dominated by an evolved stellar population \citep{Labbe2023UB,Barro2024,PerezGonzalez2024_LRD,Williams2023,Baggen2024}. Despite spectroscopic follow-up of some sources showing clear evidence for a Balmer break (which could imply that stellar continuum dominates in the rest-optical), it is not clear how such a break and broad lines can co-exist \citep{Wang2024_UB, Ma2024_LRD}. The lack of strong X-ray detections \citep{Furtak2024,Maiolino2024,Yue2024, Akins2024,Ananna2024} and apparent lack of variability \citep{Kokubo2024}, combined with the lack of hot dust emission out to rest-frame 3-4$\micron$ \citep{Williams2023,Wang2024_BRD, Akins2024} also seems to point to a stellar origin for the rest-optical Little Red Dot continuum. However, it is challenging for the stellar-only model to account for the presence of the highly luminous broad Balmer emission, especially given that the non-detection of these sources in the FIR seems to disfavor buried star formation \citep[e.g.,][]{Labbe2023LRD,Williams2023,Akins2024}. Thus, at this moment, there is no consensus on the nature of the Little Red Dots. 

Most of the works referenced above have modeled the photometric colors of Little Red Dots. Spectroscopy may provide additional important insight into the continuum shape of these systems, particularly with regard to the ``V-shaped" change in slope that Little Red Dots are selected to have. The RUBIES program with JWST/NIRSpec \citep[JWST-GO \#4233, PIs A. de Graaff and G. Brammer, see][]{deGraaff2024_RUBIES} has systematically targeted the reddest distant objects in the UDS and EGS fields with continuum emission in NIRSpec/PRISM spectroscopy at $1-5\,\mu$m. In conjunction with the many other programs that observed Little Red Dots to similar or greater depths \citep[with public spectra now hosted at the Dawn JWST Archive\footnote{\href{https://s3.amazonaws.com/msaexp-nirspec/extractions/nirspec\_graded\_v3.html}{https://s3.amazonaws.com/msaexp-nirspec/extractions/nirspec\_graded\_v3.html}} (DJA),][]{deGraaff2024_RUBIES}, it is now possible to systematically study the continuum shape of Little Red Dots in the rest-UV and -optical.

In this work, we utilize PRISM spectroscopy to study the continuum shape of Little Red Dots in detail, with the primary goal of determining where the pivot of the ``V-shape" occurs. In Section \ref{sec:data}, we describe the spectroscopic and imaging sample that we use for this work. In Section \ref{sec:analysis}, we describe our methods for identifying extremely red H$\alpha$ emitting sources and our continuum fitting method to determine the inflection location in the sources. In Section \ref{sec:break}, we discuss a range of physical models and their ability to replicate the observed Little Red Dot shape we identify. Finally, in Section \ref{sec:conclusions}, we discuss how our findings about the rest-optical spectral shape relate to studies of the full Little Red Dot spectral energy distribution (SED). Throughout this work, we adopt the best-fit cosmological parameters from the WMAP 9 year results \citep{Hinshaw2013}: $H_0 = 69.32 \ \mathrm{km \ s^{-1} \ Mpc^{-1}}$, $\Omega_m = 0.2865$, and $\Omega_\Lambda = 0.7135$, utilize a Chabrier initial mass function \citep{Chabrier2003}, and quote AB magnitudes.

\section{Data} \label{sec:data}

\subsection{JWST/NIRSpec Spectroscopy}

We obtain low-resolution ($R\sim100$) JWST/NIRSpec PRISM spectra from version 3 of the DAWN JWST Archive (DJA). These data were reduced using the latest version of \texttt{msaexp} \citep{Brammer2022}, described in detail in \citet{deGraaff2024_RUBIES}. There are two different background subtraction strategies for version 3 of the DJA; for RUBIES, we use spectra reduced with local background subtraction from nodded exposures, whereas for other programs, we use the default reductions in the DJA that were informed by the primary science goals of the programs. We note that we do not apply any additional slit loss corrections to the spectra beyond those implemented in \texttt{msaexp}. 

Our primary sample comes from the Red Unknowns: Bright Infrared Extragalactic Survey (RUBIES; GO-4233; PIs: A. de Graaff and G. Brammer). The highest priority targets in RUBIES are bright ($m_{\rm F444W}<27$) and red ($m_{\rm F150W} - m_{\rm F444W}>2$), which naturally yields a large sample of Little Red Dot-like colors. The full observing strategy is described in detail in \citet{deGraaff2024_RUBIES}.

We supplement the RUBIES sample with all public PRISM spectra available in the DJA.  
For both datasets, we select all spectra with a robust spectroscopic redshift $z>2$ (\texttt{grade=3} from visual inspection in the DJA), which ensures adequate rest-UV coverage to study the continuum of these systems, and median signal-to-noise-per-pixel $>1$. We additionally impose a magnitude limit of 26.5 in F444W imaging (see the following section). This yields 636 sources from RUBIES and 1655 sources from a variety of other programs. Our primary analysis includes spectra from the following programs: JADES \citep[JWST-GTO \#1180, PI: D. Eisenstein; JWST-GTO \#1181, PI: D. Eistenstein; JWST-GTO \#1210, PI: N. Luetzgendorf,][]{Eisenstein2023_JADES, Bunker2024_JADES}, WIDE \citep[JWST-GTO \#1211, PI: K. Isaak,][]{Maseda2024_WIDE}, JWST-GO \#1433 \citep[PI: D. Coe,][]{Meena2023}, JWST-GO \#2198 \citep[PI: L. Barrufet,][]{Barrufet2024}, UNCOVER \citep[JWST-GO \#2561, PIs: Labbe and Bezanson,][]{Bezanson2022b, Price2024}, JWST-GO \#2565 \citep[PI: K. Glazebrook,][]{Glazebrook2024, Nanayakkara2024}, JWST-DDT \#2750 \citep[PI: Arrabal Haro,][]{ArrabalHaro2023}, and JWST-DDT \#6858 (PI: D. Coulter).

\subsection{F444W Imaging}

We obtain NIRCam/F444W cutouts for all galaxies with public JWST/NIRCam imaging by querying the public NIRCam mosaics hosted on the DJA, reduced with \texttt{grizli}\citep{Brammer2023_grizli}\footnote{\href{https://grizli-cutout.herokuapp.com/}{https://grizli-cutout.herokuapp.com/}} as described in detail in \citet{Valentino2023}. We measure the {F444W} fluxes in a 0.3\arcsec aperture centered at the RA and Dec of the sources as specified in the DJA. We will also utilize these images to quantify whether these sources are extended. 

\begin{figure*}
    \centering
    \includegraphics[width=0.66\textwidth]{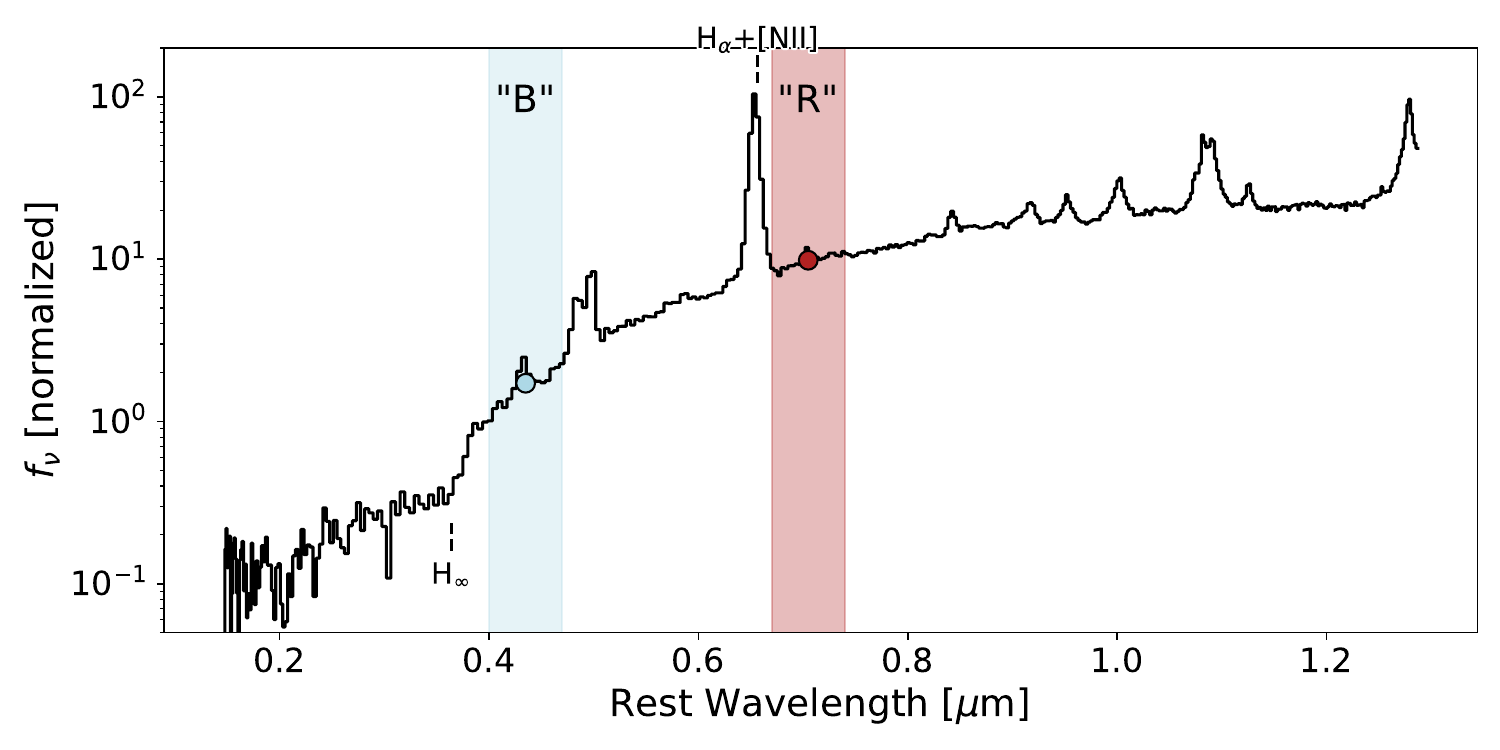}
    \includegraphics[width=0.33\textwidth]{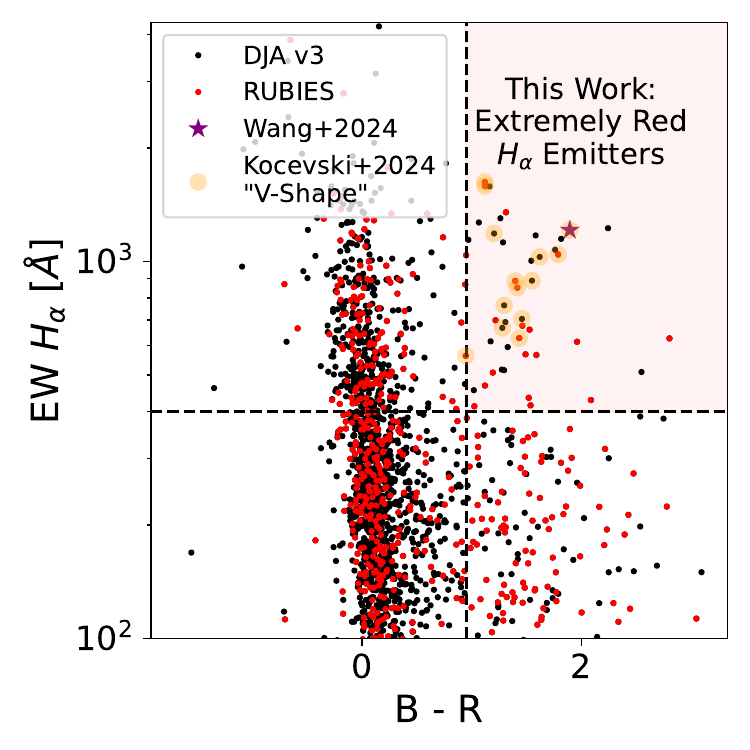}
    \caption{A demonstration of our color-EW selection of galaxies with little red dot-like rest frame colors. (Left): The spectrum of RUBIES-BLAGN-1 \citep{Wang2024_BRD}, with synthetic rest-frame ``B" and ``R" filters (chosen to avoid strong emission lines) labeled as blue and red respectively. The Balmer limit, H$_\infty$, and $H_\alpha$+[NII] are labeled. (Right): B-R versus EW H$\alpha$ for the entire RUBIES/DJA PRISM sample with $m_\mathrm{F444W}<26.5$. Sources from the RUBIES Cycle 2 program are shown as red, and all other sources from the DJA are shown as black. The top right box contains extremely red sources with strong H$\alpha$ emission, capturing sources that were selected photometrically to have a ``V-shape" in \cite{Kocevski2024} (orange), and the RUBIES-BLAGN-1 (highlighted as a purple star).}
    \label{fig:selection}

\end{figure*}

\begin{figure*}
    \centering
    \includegraphics[width=\textwidth]{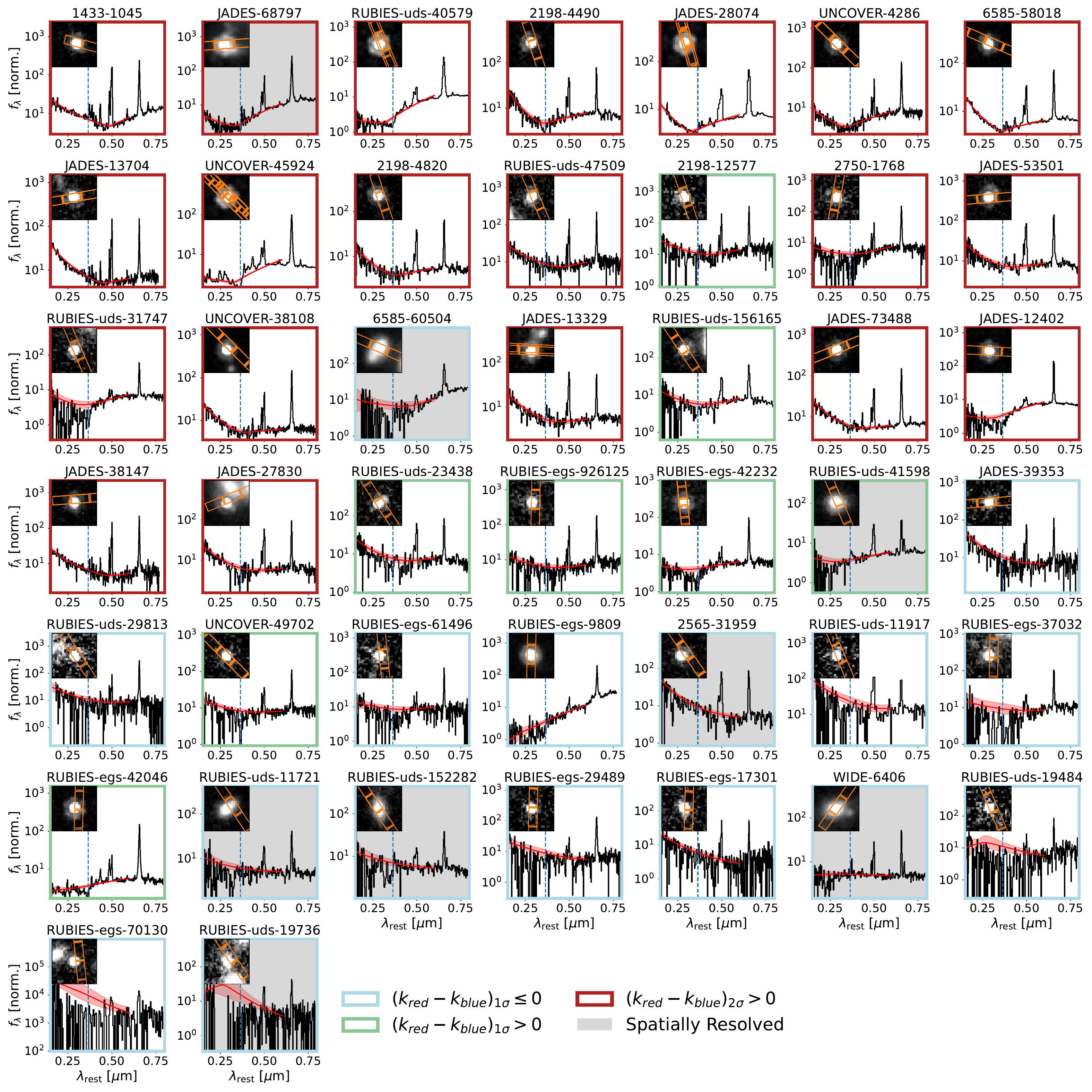}
    \caption{PRISM spectra of red, H$\alpha$ strong galaxies from the Dawn JWST Archive are shown in black, with broken power law fits to the emission-masked spectra shown in red. The spectra are ordered from top to bottom by the 50th percentile constraint on the difference between the red and blue power law indices, from reddest to bluest. Sources are outlined based on the constraints on $k_\mathrm{red}-k_\mathrm{blue}$, with $>0$ at the 2-sigma level shown in red, $>0$ at the 1-sigma level shown in green, and $\leq0$ at at the 1$\sigma$ level shown in blue. Sources with a grey background have $f(0.2"/0.1")>1.8$, which we consider to be resolved. The sources which favor a strong change in slope all change slope at an extremely similar wavelength, close to 0.3645 $\mu$m (H$_\infty$).}
    \label{fig:gallery}
\end{figure*}

\section{Analysis} \label{sec:analysis}

In this work, we seek to broadly study the continuum shape of Little Red Dots. A number of photometric Little Red Dot selection methods have been proposed in the literature, ranging from those using photometric colors \citep{Labbe2023LRD,Greene2024, Kokorev2024_LRD,Akins2024} to slopes measured from the photometry \citep{Kocevski2024}. Here, we step back from any cuts that enforce the presence of a ``V-shape" or compactness and instead employ a simple set of cuts to select a pure red sample that is emitting strongly in H$\alpha$. From this sample, we then apply a number of further cuts (e.g., a determination of whether or not the source has a ``V-shape" or is compact, meeting the definition of a Little Red Dot) to determine how common Little Red Dots are among this extremely red sample and what features in the continuum shape these Little Red Dots share in common. We note that one angle that we do not explore in depth is the width of H$\alpha$--this is predominantly due to the resolution limit of the PRISM spectra, which typically have R$\sim100$ at the wavelength of H$\alpha$ in this sample. Future work will fold in this information about the width of the emission for samples where higher-resolution spectra exist.

\subsection{Extremely red H$\alpha$ emitter selection}

Here, we outline our selection of extremely red $H\alpha$ emitters with a simple rest-frame color and equivalent width selection. Selecting on rest-frame colors has the advantage of ensuring that we are tracing the continuum shape without significant emission line boosting, while still selecting for galaxies with strong H$\alpha$ via an equivalent width cut allows us to select for likely broad sources without requiring a determination of the line width at PRISM resolution for low-SNR sources. These cuts are deliberately broad and will by definition include resolved extremely red galaxies \citep[e.g.,][]{Gentile2024} and more ``typical" reddened quasars \citep[e.g.,][]{Assef2016} in addition to Little Red Dots; we will then whittle this sample down to quantify what fraction of these sources meet Little Red Dot criteria.

First, we define a set of $0.07$ $\mu$m width rest-frame filters which are designed to avoid strong emission and probe the rest optical. The first filter, ``B", spans 0.40-0.47 $\mu$m. The blue end was chosen to explicitly avoid the Balmer breaks, ensuring that we are selecting for all sources that are red independent of whether or not they have a break or ``V shape". The red end was chosen to terminate just blueward of H$\beta$. The second filter ``R", spans 0.67-0.74 $\mu$m, with the blue end motivated by starting redward of H$\alpha$ and the red end set to keep the filter width identical to the B filter. We require that any galaxies we select have rest-frame wavelength coverage that spans the full range of our selection, effectively limiting us to galaxies at $z<6.15$ due to the truncation of the red filter. We demonstrate these filters overlaid on the RUBIES-BLAGN-1, a broad line emitting red source at $z=3.1$ \citep{Wang2024_BRD}, in the left panel of Figure \ref{fig:selection}.

In addition to selecting on color, we also select based on the rest-frame equivalent width of H$\alpha$+[N\,{\sc ii}]. To measure this equivalent width, we fit for a power law continuum between 0.58 and 0.73 $\mu$m, masking between 0.63 and 0.67 $\mu$m, and use this wavelength-dependent continuum to integrate the equivalent width in the same 0.63-0.67 $\mu$m range. Again, while we do not explicitly select for broad lines, a high-EW selection can serve as a proxy that ensures that our selected sources could at least potentially host the highly luminous broad H$\alpha$ that defines Little Red Dots.

Using this color and equivalent width and the magnitudes measured as specified in Section \ref{sec:data}, we select all galaxies from the DJA that meet the following criteria:

\begin{enumerate}
    \item B - R $>$ 0.95
    \item EW($\rm H\alpha) > 400$ \angstrom
    \item $m_{\rm F444W}<26.5$
\end{enumerate}

The first two cuts are shown in the right panel of Figure \ref{fig:selection}, where we show the full DJA in this color plane and highlight this selection box in red, with literature samples of Little Red Dots selected photometrically with ``V shapes" \citep{Kocevski2024} shown highlighted in orange, RUBIES-BLAGN-1 \citep{Wang2024_BRD} shown as a purple star, and DJA/RUBIES sources shown in black and red respectively. The color and equivalent width cuts were motivated to include the vast majority of literature ``V shapes" without explicitly selecting for any UV break. Because we do not explicitly select for compactness or incorporate any rest-UV information into our selection, we are effectively selecting extremely red, strong H$\alpha$ emitters. The magnitude cut is motivated  by the depth of the RUBIES PRISM spectroscopy (48 min exposure times), and corresponds to a median continuum $\mathrm{SNR}\sim3$ per pixel for a well-centered point source. In total, the selection yields 44 sources with extremely red colors and high-equivalent width $H\alpha$.

Finally, we also attempt to quantify whether the sources in our sample are resolved. To do so, we take the ratio of the flux in a 0.2$\arcsec$ circular aperture to the flux in a 0.1$\arcsec$ aperture for the NIRCam/F444W imaging. We use the empirically derived PSF models for the F444W band from the UNCOVER survey \citep[PIs: R. Bezanson and I. Labbe,][]{Bezanson2022b, Weaver2024} to estimate that this ratio should be $\sim$1.5 for unresolved sources. Accounting for the varying dithering strategies and depths of the imaging used here in addition to the centering of the slits on sources, we conservatively consider sources with $f(\frac{0.2''}{0.1''})>1.8$ to be resolved, and note that this distinction is purely qualitative and does not enter rigorously into our analysis.

A full gallery of the spectra of these extremely red H$\alpha$ emitters is shown in Figure \ref{fig:gallery}, with insets showing the F444W images of the sources. Sources with $f(\frac{0.2''}{0.1''})>1.8$ are highlighted in grey. While we did not explicitly select for compactness, the vast majority of the sources in this sample do indeed appear extremely compact. For each source, we mark H$_\infty$ (0.3645 $\mu$m) with a dashed line to illustrate that many galaxies exhibit strong inflections in their spectral slope at this specific wavelength. In the next section, we will quantify that impression.

\subsection{Determining spectral slope inflection strength and location} \label{subsec:break_location}

The primary goal of this work is to study the continuum shape of spectroscopically confirmed extremely red H$\alpha$ emitters. We aim to determine whether the finding that some specific Little Red Dot systems have strong inflections that resemble Balmer breaks \citep[e.g.,][Labbe et al. in preparation]{Wang2024_BRD, Wang2024_UB, Ma2024_LRD} is a feature of the population of these objects as a whole. As such, we require a metric that satisfies the following conditions:

\begin{enumerate}
    \item The metric can determine whether sources exhibit a sharp change in slope.
    \item The metric can, at least approximately, determine where that change in slope occurs.
    \item The metric remains agnostic to the physics of that inflection, meaning the change in slope can occur at any wavelength.
\end{enumerate}

To that end, we fit the spectra that fall into our extremely red $H\alpha$ emitting selection with a simple broken power law model taking the following form:

\begin{equation}
    f_\nu \propto \begin{cases}
        \lambda ^ {k_\mathrm{blue}} \ (\lambda < \lambda_\mathrm{break}) \\
        \lambda ^ {k_\mathrm{red}} \ (\lambda > \lambda_\mathrm{break}) \\
    \end{cases}
\end{equation}

\begin{table*}[]
    \centering
    \begin{tabular}{rrrrrrrrrrrr}
\toprule
Program & ID\footnote{For RUBIES sources, we list both the field and ID, as fields do not have unique ID numbers. For JADES sources, the ID number represents the spectroscopic, not photometric, ID} & RA & Dec & z & $m_{F444W}$ & B-R & $EW_{H\alpha} \ [$\AA$]$ & $f(\frac{0.2''}{0.1''})$ & $\lambda_\mathrm{break}$ \ [$\mu$m] & $\Delta k$ \\
    \hline
1433 & 1045 & 101.9334 & 70.1983 & 4.5295 & 24.5 & 1.81 & 1149 & 1.6 & 0.49 $\pm ^{0.01}_{0.01}$ & 4.0 $\pm ^{0.8}_{0.8}$ \\
JADES & 68797 & 189.2291 & 62.1462 & 5.0352 & 22.6 & 2.24 & 1226 & 1.84 & 0.34 $\pm ^{0.02}_{0.02}$ & 3.7 $\pm ^{0.5}_{0.5}$ \\
RUBIES & uds-40579 & 34.2442 & -5.2459 & 3.1124 & 22.3 & 1.89 & 1212 & 1.32 & 0.32 $\pm ^{0.02}_{0.02}$ & 3.3 $\pm ^{0.6}_{0.6}$ \\
2198 & 4490 & 53.041 & -27.8545 & 3.7011 & 25.4 & 1.27 & 517 & 1.47 & 0.34 $\pm ^{0.03}_{0.04}$ & 3.3 $\pm ^{0.6}_{0.6}$ \\
JADES & 28074 & 189.0646 & 62.2738 & 2.2723 & 22.1 & 1.54 & 1005 & 1.35 & 0.32 $\pm ^{0.01}_{0.01}$ & 3.2 $\pm ^{0.3}_{0.3}$ \\
UNCOVER & 4286 & 3.6192 & -30.4233 & 5.8353 & 25.1 & 1.55 & 890 & 1.36 & 0.34 $\pm ^{0.03}_{0.02}$ & 3.1 $\pm ^{0.5}_{0.5}$ \\
6585 & 58018 & 150.1444 & 2.3444 & 3.9278 & 24.3 & 1.46 & 705 & 1.38 & 0.35 $\pm ^{0.01}_{0.01}$ & 3.0 $\pm ^{0.2}_{0.2}$ \\
JADES & 13704 & 53.1265 & -27.8181 & 5.9296 & 26.2 & 1.3 & 765 & 1.43 & 0.44 $\pm ^{0.03}_{0.03}$ & 2.9 $\pm ^{0.8}_{0.6}$ \\
UNCOVER & 45924 & 3.5848 & -30.3436 & 4.4668 & 22.5 & 1.06 & 1270 & 1.33 & 0.33 $\pm ^{0.01}_{0.01}$ & 2.6 $\pm ^{0.3}_{0.2}$ \\
2198 & 4820 & 53.0349 & -27.8522 & 3.7966 & 25.5 & 1.28 & 667 & 1.53 & 0.37 $\pm ^{0.04}_{0.04}$ & 2.6 $\pm ^{0.6}_{0.6}$ \\
RUBIES & uds-47509 & 34.2646 & -5.2326 & 5.6752 & 25.0 & 1.42 & 853 & 1.5 & 0.44 $\pm ^{0.04}_{0.06}$ & 2.6 $\pm ^{1.1}_{1.0}$ \\
2198 & 12577 & 53.0485 & -27.8151 & 5.232 & 25.5 & 1.76 & 1075 & 1.32 & 0.42 $\pm ^{0.05}_{0.07}$ & 2.5 $\pm ^{1.3}_{1.2}$ \\
2750 & 1768 & 214.9258 & 52.9457 & 5.0946 & 26.1 & 1.62 & 1029 & 1.35 & 0.4 $\pm ^{0.06}_{0.07}$ & 2.4 $\pm ^{1.0}_{1.1}$ \\
JADES & 53501 & 189.2951 & 62.1936 & 3.4283 & 25.0 & 1.29 & 1124 & 1.35 & 0.4 $\pm ^{0.06}_{0.07}$ & 2.4 $\pm ^{1.0}_{0.9}$ \\
RUBIES & uds-31747 & 34.2238 & -5.2602 & 4.1305 & 25.1 & 1.54 & 416 & 1.38 & 0.36 $\pm ^{0.07}_{0.05}$ & 2.4 $\pm ^{1.0}_{1.1}$ \\
UNCOVER & 38108 & 3.53 & -30.358 & 4.962 & 25.0 & 1.2 & 1187 & 1.34 & 0.37 $\pm ^{0.04}_{0.04}$ & 2.3 $\pm ^{0.5}_{0.5}$ \\
6585 & 60504 & 150.1432 & 2.356 & 2.9876 & 22.9 & 2.55 & 509 & 2.55 & 0.44 $\pm ^{0.05}_{0.11}$ & 2.3 $\pm ^{2.3}_{2.7}$ \\
JADES & 13329 & 53.139 & -27.7844 & 3.9441 & 25.5 & 1.29 & 515 & 1.52 & 0.41 $\pm ^{0.06}_{0.07}$ & 2.2 $\pm ^{1.0}_{0.9}$ \\
RUBIES & uds-156165 & 34.5005 & -5.1273 & 2.7685 & 24.6 & 1.53 & 660 & 1.33 & 0.4 $\pm ^{0.06}_{0.06}$ & 2.2 $\pm ^{1.2}_{1.2}$ \\
JADES & 73488 & 189.1974 & 62.1772 & 4.1304 & 25.3 & 1.26 & 1302 & 1.35 & 0.4 $\pm ^{0.04}_{0.04}$ & 2.2 $\pm ^{0.6}_{0.5}$ \\
JADES & 12402 & 53.1327 & -27.7655 & 3.1965 & 24.3 & 1.58 & 1173 & 1.38 & 0.32 $\pm ^{0.04}_{0.03}$ & 2.1 $\pm ^{0.9}_{0.9}$ \\
JADES & 38147 & 189.2707 & 62.1484 & 5.8778 & 25.2 & 1.16 & 1582 & 1.4 & 0.47 $\pm ^{0.02}_{0.06}$ & 2.1 $\pm ^{1.4}_{1.2}$ \\
JADES & 27830 & 189.2545 & 62.2493 & 3.9386 & 24.5 & 1.31 & 691 & 1.68 & 0.4 $\pm ^{0.06}_{0.06}$ & 2.1 $\pm ^{1.0}_{1.0}$ \\
RUBIES & uds-23438 & 34.3383 & -5.2809 & 3.6907 & 24.9 & 1.46 & 676 & 1.52 & 0.42 $\pm ^{0.06}_{0.08}$ & 2.0 $\pm ^{1.5}_{1.3}$ \\
RUBIES & egs-926125 & 215.1371 & 52.9886 & 5.284 & 25.5 & 1.79 & 1044 & 1.34 & 0.42 $\pm ^{0.06}_{0.09}$ & 1.6 $\pm ^{1.5}_{1.5}$ \\
RUBIES & egs-42232 & 214.8868 & 52.8554 & 4.965 & 24.9 & 1.4 & 888 & 1.36 & 0.34 $\pm ^{0.08}_{0.05}$ & 1.4 $\pm ^{1.0}_{1.0}$ \\
RUBIES & uds-41598 & 34.2305 & -5.244 & 3.1934 & 23.1 & 1.52 & 435 & 1.94 & 0.34 $\pm ^{0.09}_{0.06}$ & 1.3 $\pm ^{0.9}_{0.9}$ \\
JADES & 39353 & 189.294 & 62.1531 & 4.851 & 26.4 & 1.17 & 614 & 1.32 & 0.42 $\pm ^{0.06}_{0.09}$ & 1.3 $\pm ^{1.6}_{1.3}$ \\
RUBIES & uds-29813 & 34.4534 & -5.2707 & 5.4435 & 25.9 & 1.12 & 1625 & 1.37 & 0.41 $\pm ^{0.06}_{0.11}$ & 1.2 $\pm ^{2.1}_{2.2}$ \\
UNCOVER & 49702 & 3.5477 & -30.3337 & 4.8756 & 26.4 & 1.33 & 593 & 1.31 & 0.39 $\pm ^{0.07}_{0.08}$ & 1.2 $\pm ^{1.0}_{1.0}$ \\
RUBIES & egs-61496 & 214.9724 & 52.9622 & 5.0764 & 26.1 & 1.43 & 627 & 1.38 & 0.39 $\pm ^{0.07}_{0.09}$ & 1.2 $\pm ^{1.4}_{1.5}$ \\
RUBIES & egs-9809 & 215.0173 & 52.8802 & 5.6844 & 23.5 & 2.8 & 626 & 1.56 & 0.37 $\pm ^{0.07}_{0.07}$ & 1.0 $\pm ^{1.2}_{1.2}$ \\
2565 & 31959 & 34.4836 & -5.1689 & 3.4674 & 24.9 & 0.97 & 1142 & 1.82 & 0.42 $\pm ^{0.06}_{0.1}$ & 0.9 $\pm ^{1.4}_{1.2}$ \\
RUBIES & uds-11917 & 34.2683 & -5.2952 & 2.5909 & 26.1 & 1.49 & 568 & 1.32 & 0.41 $\pm ^{0.07}_{0.12}$ & 0.7 $\pm ^{3.0}_{2.6}$ \\
RUBIES & egs-37032 & 214.8494 & 52.8118 & 3.8467 & 25.4 & 1.96 & 613 & 1.4 & 0.39 $\pm ^{0.08}_{0.1}$ & 0.6 $\pm ^{2.4}_{2.5}$ \\
RUBIES & egs-42046 & 214.7954 & 52.7888 & 5.2702 & 23.5 & 1.12 & 1587 & 1.34 & 0.33 $\pm ^{0.09}_{0.05}$ & 0.6 $\pm ^{0.6}_{0.6}$ \\
RUBIES & uds-11721 & 34.411 & -5.3008 & 3.9807 & 24.2 & 1.12 & 469 & 1.91 & 0.38 $\pm ^{0.09}_{0.09}$ & 0.3 $\pm ^{1.2}_{1.3}$ \\
RUBIES & uds-152282 & 34.506 & -5.1327 & 3.429 & 24.1 & 1.02 & 413 & 1.98 & 0.39 $\pm ^{0.08}_{0.09}$ & 0.3 $\pm ^{1.4}_{1.4}$ \\
RUBIES & egs-29489 & 215.0221 & 52.9208 & 4.5423 & 26.1 & 1.31 & 1351 & 1.28 & 0.39 $\pm ^{0.08}_{0.11}$ & 0.2 $\pm ^{2.0}_{2.0}$ \\
RUBIES & egs-17301 & 214.9875 & 52.8731 & 5.2246 & 26.1 & 1.22 & 699 & 1.56 & 0.4 $\pm ^{0.07}_{0.11}$ & -0.2 $\pm ^{2.0}_{1.7}$ \\
WIDE & 6406 & 189.2761 & 62.2142 & 4.4108 & 24.2 & 1.02 & 456 & 2.29 & 0.36 $\pm ^{0.1}_{0.08}$ & -0.4 $\pm ^{0.7}_{0.8}$ \\
RUBIES & uds-19484 & 34.2324 & -5.2807 & 4.6627 & 26.2 & 2.09 & 429 & 1.46 & 0.32 $\pm ^{0.13}_{0.06}$ & -1.8 $\pm ^{3.0}_{3.1}$ \\
RUBIES & egs-70130 & 214.9508 & 52.9669 & 2.9766 & 25.5 & 1.59 & 565 & 1.36 & 0.35 $\pm ^{0.11}_{0.08}$ & -2.4 $\pm ^{4.3}_{4.4}$ \\
RUBIES & uds-19736 & 34.3011 & -5.2878 & 4.8024 & 25.4 & 1.19 & 508 & 2.0 & 0.28 $\pm ^{0.14}_{0.02}$ & -4.4 $\pm ^{4.1}_{4.1}$ \\
\hline
    \end{tabular}
    
    \caption{A table of the all extremely red (B-R$>0.95$) H$\alpha$ emitters (rest-frame EW$>400 \ \AA$) from the RUBIES survey or public PRISM data. Sources are sorted by their median $k_\mathrm{red}-k_\mathrm{blue}$ ($\Delta k$), as determined by the fits described in Section \ref{subsec:break_location}}.
    \label{tab:LRD_sample}
\end{table*}

This model is quite similar to the rolling bandpass method employed by \cite{Kocevski2024} in identifying Little Red Dots from photometry, making it a valuable tool for flagging galaxies where such a ``V shape" occurs, with the additional feature of placing some constraints on the location of that inflection point by not enforcing that the inflection occurs at $H_\infty$. We use \texttt{emcee} \citep{Foreman-Mackey2013} to fit this model to all of our extremely red H$\alpha$ emitters in the range $0.15\,\mathrm{\mu m} < \lambda < 0.6$ $\mu$m (deliberately choosing a wide wavelength range to remain agnostic to the specific inflection location), with the normalization ($a>0$), the break wavelength ($\lambda_\mathrm{break}\in [0.25, 0.5]$), and two power law indices ($k_\mathrm{blue}$ and $k_\mathrm{red} \in [-10,10]$) as free parameters. In order to account for model mismatch between this simple broken power law, the true shape of these systems, and the flux calibration being uncertain, we apply ``grease" to the models in two ways. The first is enforcing a floor in the signal-to-noise of 20. Second, we fit for a free white noise error inflation term ($\in [0.1,5]$) that is penalized in our likelihood function. We mask the 0.005 $\mu$m region on either side of all strong emission features in this wavelength range, including [O\,{\sc ii}], H$\delta$, H$\gamma$, and He\,{\sc i} in these fits, and evaluate likelihood in $\log(f_\nu)$ when comparing to the model, effectively masking all negative fluxes.

In Figure \ref{fig:gallery}, we show the median (red line) and 1$\sigma$ (shaded region) constraints for all our fits. The gallery is sorted in descending order of median constraints on $k_\mathrm{red} - k_\mathrm{blue}$ (which we refer to as the inflection strength). Sources which are consistent at the $2\sigma$ level with $k_\mathrm{red} - k_\mathrm{blue}>0$ have frames outlined in red, sources which are consistent with $k_\mathrm{red} - k_\mathrm{blue}>0$ at the $1\sigma$ level have frames outlined in green, and all other sources have frames outlined in blue. In each panel, we mark $0.3645$ $\mu$m, the limit of the Balmer series, as a dashed line for illustrative purposes. Visual inspection of the well-determined red inflection sample suggests that the majority, if not all, of the strong changes in slope occur quite near to this wavelength. {However, it is striking that despite this, there is a wide variety of shapes of these changes in slope, ranging from transitions where a weak UV suddenly jumps (e.g., UNCOVER-45924, see also Labbe et al. in preparation) to strong UV components that quickly change slopes without necessarily ``breaking" (e.g., JADES-28074, 6585-58018). We also note that there appears to be a distinction between the concavity of the red continuum in these two classes of objects, with the former tending to become shallower redward of the inflection point and the latter better resembling an already-shallow power law. The results of this fitting are also shown in Table \ref{tab:LRD_sample} in the same order (left to right) as they appear in Figure \ref{fig:gallery}.

In order to explore this quantitatively, in Figure \ref{fig:break-deltak} we plot the inflection strength versus the wavelength of the break, with our well determined ($>2\sigma$) sample highlighted as red points. We additionally highlight as solid sources where the break location is considered to be ``well-determined", with $\lambda_\mathrm{break,84}-\lambda_\mathrm{break,16}<750 \AA$. However, given that the double power law is a poor descriptor of the actual shape of some of these sharp breaks (see for example RUBIES-uds-40579 and UNCOVER-45924), it is not surprising that the break location we measure does not perfectly align with H$_\infty$. To test this, we fit our double power law model to a range of galaxy-only models (drawn from a grid of $t_\mathrm{age}=[0.1,0.4]$ Gyr, $\tau_V=[2-4]$ (assuming a \citealt{Kriek2013} dust law with a steep dust index of -1, see \citealt{Ma2024_LRD}), and $f_\mathrm{scatter}=[0.5\%-2\%]$, see Section \ref{subsec:galaxy-galaxy}) with strong Balmer breaks, and find that fits to breaks that are purely stellar in origin span a range of $2.5\gtrsim k_\mathrm{red} - k_\mathrm{blue} \gtrsim 6$ and $0.3 \gtrsim \lambda_\mathrm{break} \gtrsim 0.35$. Visual inspection of the population with strong changes in spectral slope confirms that the location of the inflection point itself is quite uniform. We acknowledge two sources with well-determined breaks that do not fall into this region; in both these sources (1433-1045 and JADES-13704), there is strong Balmer line emission and a feature that resembles a Balmer jump \citep[e.g.,][]{Katz2024} directly before the slope becomes red, suggesting that these sources also seem to share the clear association with a transition associated with the Balmer series.

Finally, we note that the sample where no inflection location is determined is highly heterogeneous. Visual inspection of these sources reveals that some may indeed have changes in slope that are washed out by the low continuum signal-to-noise (e.g., RUBIES-egs-37032); many of these sources also appear to have slits that are poorly aligned with sources, making recovery of their rest-frame UV slopes difficult. Others, mostly among the resolved sample, clearly resemble dusty star forming galaxies. It is noteworthy that thse are among the only sources that appear to have prominent [SII] emission, suggesting a significanty different ionizing mechanism (and potentially metallicity) to the unresolved ``V shaped" sources. And some number of these sources appear to be completely unresolved and have extremely red shapes but lack a strong UV component, resembling typical reddened quasars \citep[e.g.,][]{Glikman2012,Banerji2015}.

Sources with strong red changes in slope are essentially all unresolved in their F444W imaging; only one source, JADES-68797, is considered resolved by our compactness criteria due to the presence of a diffuse component. However, the core of this source is still clearly compact as evidenced by the diffraction spikes. As such, these sources meet the ``V-shaped", compact Little Red Dot criteria employed in the literature \citep[e.g.,][]{Labbe2023LRD, Greene2024, Kokorev2023_LRD, Kocevski2024}, and we make the empirical statement that sources in our sample with well-determined changes in spectral slope are ``V shaped" Little Red Dots, and that Little Red Dot inflection points consistently occur at $3645$ \angstromcomma, the Balmer limit.

\begin{figure}
    \centering
    \includegraphics[width=0.5\textwidth]{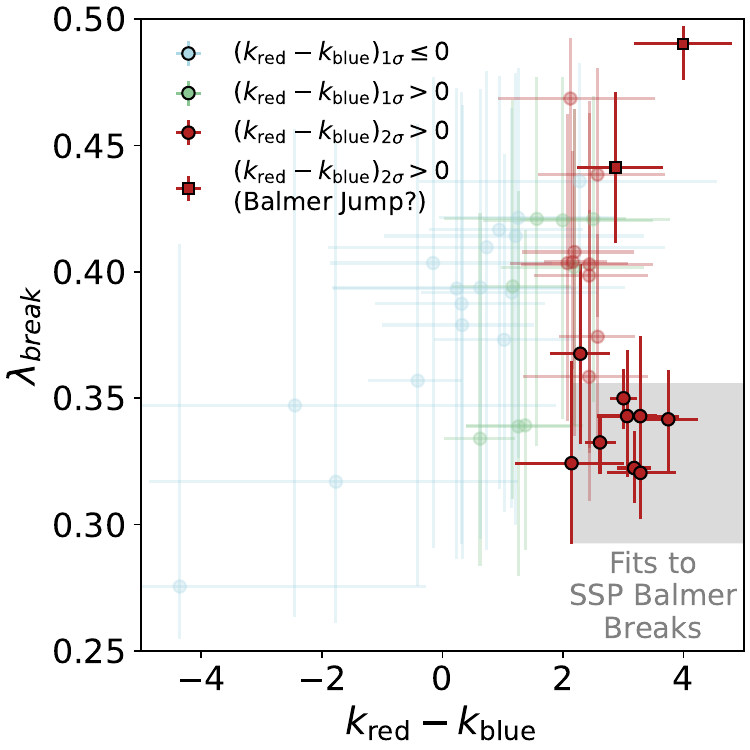}
    \caption{The results of the broken power law fits to the red spectra, illustrated by showing the break wavelength versus the difference between the red and blue power law indices. Sources with $k_\mathrm{red} - k_\mathrm{blue}>0$ at the $2\sigma$ level are colored red, those consistent with $k_\mathrm{red} - k_\mathrm{blue}>0$ at the $1\sigma$ level are colored green, and those consistent no break are colored blue. Sources which have a well determined break ($\lambda_\mathrm{break, 84} - \lambda_\mathrm{break, 16} < 750 \AA$) are shown as solid points, and all other sources are shown with transparency. The shaded grey region shows the range of $k_\mathrm{red} - k_\mathrm{blue}$ and $\lambda_\mathrm{break}$ that comes from applying this model to PRISM models of dusty post-starburst galaxies (with some scattered light, see Section \ref{subsec:galaxy-galaxy}). All of the galaxies with a preference for a strong change in slope are fit with a break that occurs near $0.3645$ $\mu$m, the Balmer break, at a very similar wavelength to the one returned by model fits to simple stellar populations.}
    \label{fig:break-deltak}
\end{figure}

\begin{figure*}
    \centering
    \includegraphics[width=\textwidth]{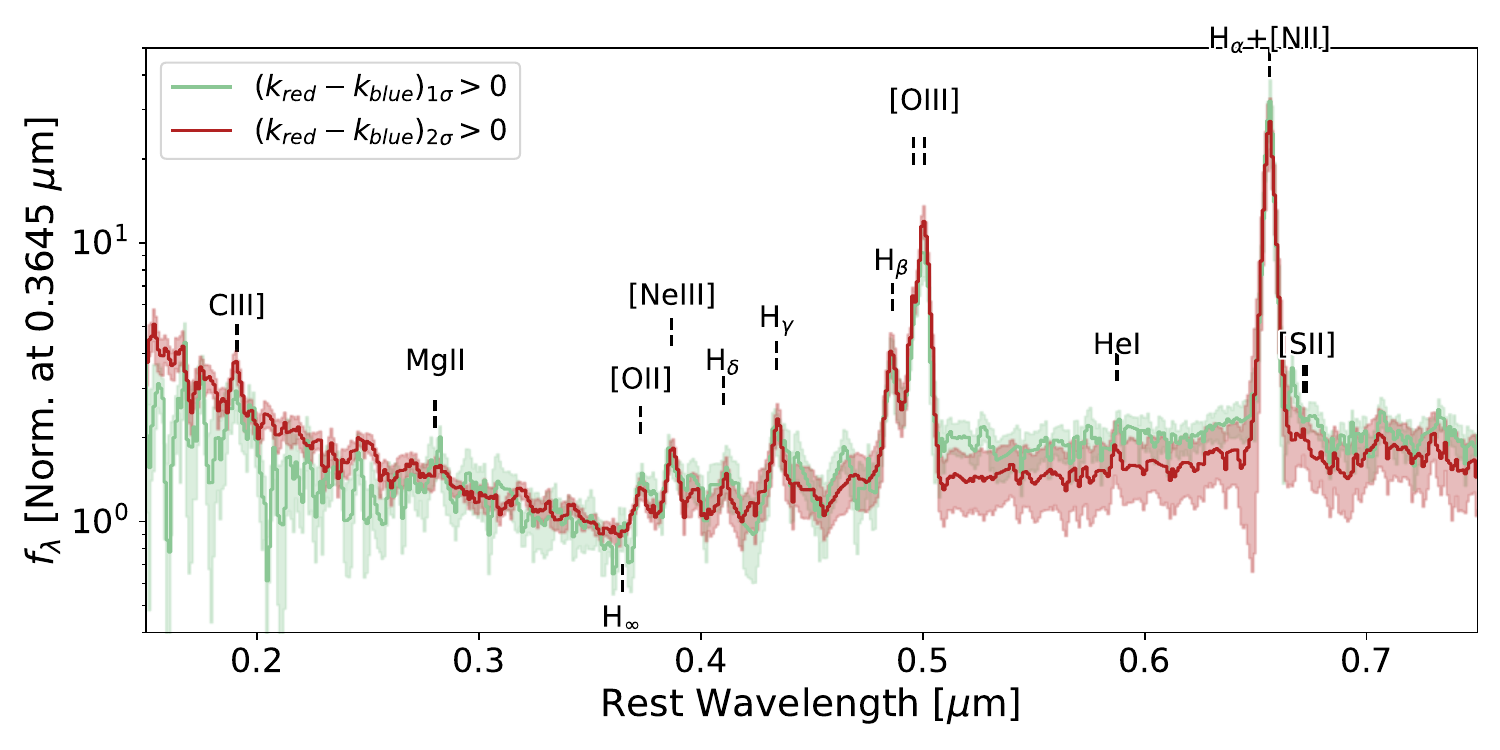}
    \caption{Median stacks of the $2\sigma$ (red) and $1\sigma$ (green) samples that are unresolved (see Fig \ref{fig:break-deltak}). Shaded regions denote the error in the median. The sources well constrained red changes in slope show a strong preference for a inflection at and around $0.3645$ $\mu$m, H$_\infty$. The stack of sources with less well constrained changes in slope still exhibit a similar shape at H$_\infty$. Key emission features are labeled.}
    \label{fig:two_stack}
\end{figure*}

\section{Why are changes in spectral slopes among Little Red Dots associated with the Balmer limit?} \label{sec:break}

In the previous section, we demonstrated that a significant portion (20/44) of high H$\alpha$ EW extremely red galaxies have strong changes in spectral slope (measured at the $2\sigma$ level) that consistently occur around $\sim3645$ \angstromcomma. This likely represents a lower limit on the true fraction of extremely red H$\alpha$ emitters with such inflection points, as our stringent criteria for a significant change in slope would make weak changes hard to detect in low signal-to-noise sources or sources with fainter intrinsic UV continuum--many of these sources appear in our $1\sigma$ sample. Models have explained this inflection point using a variety of prescriptions, including a reddened quasar with a UV dominated by scattered light \citep[e.g.,][]{Labbe2023LRD, Furtak2024, Barro2024, Greene2024, Akins2024}, a reddened AGN with a UV dominated by stellar continuum \citep[e.g.,][]{Wang2024_BRD, Wang2024_UB, Greene2024, Ma2024_LRD}, a continuum fully dominated by stellar light \citep[e.g,][]{Akins2023, Labbe2023UB, Williams2023}, or absorption from a non-stellar source of n=2 hydrogen \citep{Inayoshi2024}. In this section, we use a set of toy models to test whether these scenarios are at odds with our finding that the inflection point in the ``V shape" occurs at a consistent wavelength.

In order to do so, we construct an over-sampled stack of the 19 sources that are fit with strong changes in slope (excluding the one resolved source with a well-determined inflection point). We interpolate each rest-frame spectrum onto a grid spanning 0.2 $\mu$m to 0.7 $\mu$m in bins of 0.001 $\mu$m after normalizing the median flux density in the 100 \angstrom window around H$_\infty$ to be 1. We then take the median flux of all sources in each super-sampled pixel, weighting each galaxy equally, and we also measure the standard error in the median of each of these pixels.

We show this stack in Figure \ref{fig:two_stack} in red, showing the standard error in the median as a shaded region. For illustrative purposes, we also show the stack of the unresolved sources that are consistent with changes in slope at the $1\sigma$ level in green. The stack of Little Red Dots shows an extremely strong preference for an abrupt jump right at H$_\infty$, with a characteristically blue slope blueward of the Balmer limit and a wider range of red end slopes. The $1\sigma$ stack appears similar, indicating again that a significant fraction of the sources in that sample have common inflection points. Both stacks show strong Balmer series emission, in addition to C\,{\sc iii}], [O\,{\sc ii}], [O\,{\sc iii}], [NeIII], and He\,{\sc i}. 

In this work, we are primarily concerned with the physical origin of the strong changes in slope among Little Red Dots. Based on our analysis, it is clear that any model that can explain these spectral inflection points must consistently produce a sharp change in slope at H$_\infty$ without invoking an extremely specific combination of reddening and fractional contribution of two components and result in an SED that is red enough to fall into our B-R color selection,

In the following sections, we employ a set of toy models to assess whether Little Red Dot models proposed in the literature can explain the sharp transitions seen in a large fraction of extremely red galaxies. We stress that we are not fitting any models in this section to the stacked data. Instead, we use the models to illustrate what types of models can satisfy the above criteria by comparing to the shape of the stack at and around 3645 \angstromcomma. As a general rule, any model that is plotted with a solid line meets the $B-R>0.95$ criteria for ``extremely red" outlined in Section \ref{sec:analysis}, while models that do not are plotted as dashed lines. All stellar population models are generated with Flexible Stellar Population Synthesis \citep{Conroy2009, Conroy2010} with nebular emission turned off, and all AGN models are generated using the \cite{Temple2021} model.

\subsection{An inflection arising from the transition between reddened AGN continuum and scattered light} \label{subsec:agn-only}

Because of their broad lines and compact morphologies, it has been suggested that Little Red Dots are hosts to obscured AGN, with their continuum shape dominated by a reddened accretion disk \citep[e.g.,][]{Labbe2023LRD, Furtak2024, Barro2024,Greene2024}. In this model, the ``V-shape" of the spectrum results from an intersection between a reddened AGN continuum and a significantly smaller scattered light component that has the spectrum of the unreddened AGN \citep{Furtak2023,Greene2024}. Many dusty quasars, both in the local universe \citep{Veilleux2013} and at high-redshift \citep{Glikman2012,Assef2016, Hamann2017} are well described by such a model.

To model the two-component AGN-only model, we follow \cite{Akins2024} and \cite{Ma2024_LRD} and assume that the intrinsic AGN spectral energy distribution follows the empirically measured \cite{Temple2021} continuum shape derived from type 1 quasars at $0<z<5$. We assume that some fraction of this light is attenuated following a \cite{Kriek2013} dust law, with a dust index that controls the offset of the curve relative to the \cite{Calzetti1997} dust law and parameterizes the UV bump, $\delta$, and a normalization $\tau_V$, and that the rest of the light is not attenuated. 

In Figure \ref{fig:model_test}a, we show the stack of our Little Red Dot sample with strong red changes in slope in red, along with a set of models generated using one intrinsic AGN model under a range of attenuation assumptions. All models are set to $f_{scatter} = 1\%$ and are normalized so that the scattered light component is equal to our stack at $3645$ \angstromcomma. This scattered light component is shown as a dotted black line. We then show a range of 4 AGN attenuation models, with the remaining 99\% of the light subjected to dust screens of $\tau_V=$ 1, 2, 3, and 4. For all these models, we set the dust index (see \citealt{Kriek2013}) to $-1$, which is steeper than any commonly used dust law \citep[for more discussion, see][]{Ma2024_LRD}.

Even under the assumption of an extremely steep dust law, this model does not prove to be a good description of the observed spectral shape of Little Red Dots. First, it is difficult for the shape of the inflection to be steep enough to appear as an abrupt jump in the spectrum; this is because the intersection of the scattered and extincted AGN power laws results in a smooth transition between the two components. But perhaps even more importantly, this model fails to meet our criteria for describing the shape of Little Red Dots because there is no reason that these weak changes in slope should consistently occur at some preferred wavelength. Indeed, the models that would result in a system that is red enough to meet our B-R selection \citep[$\tau_V=$ 2, 3, and 4, similar to the range of attenuation found to result in observed Little Red Dot colors in][]{Volonteri2024} show inflection points that range from rest-frame $\sim$0.25 $\mu$m out to $\sim$0.4 $\mu$m. And while for any given $\tau_V$, there exists a scattering fraction that could conspire to locate the inflection right around 3645 \angstromcomma, it would be extremely unlikely for all observed systems to be configured precisely such that this change in slope occurs at the same location. We note that this claim is not unique to the assumed \cite{Temple2021} AGN model; any AGN model with a smooth continuum that does not feature any intrinsic inflection points \citep[e.g.,][]{Kubota2018} would fail to replicate the observed Little Red Dot spectral shape.

Thus, we claim that it is extremely unlikely that the typical spectrum of Little Red Dots with strong changes in spectral slope can be the result of a two-component AGN model that produces this change via the intersection of a scattered and an attenuated component. Even putting aside the issue of the steepness of the red end slope \citep[which could result from some different intrinsic AGN shape or dust grain size distributions that result in an even steeper dust law, see][]{Ma2024_LRD} and the issues related to the deficiency of X-ray \citep{Furtak2024, Maiolino2024, Yue2024, Akins2024, Ananna2024} and rest-frame 3-4$\mu$m emission \citep{Williams2023, Akins2024, Wang2024_BRD} emission typically seen in AGN, the consistency and sharpness of the inflection point is highly unlikely to result from any model that invokes the intersection of two components with no intrinsic changes in slope to explain the observed spectral shapes. 

\begin{figure*}
    \centering
    \includegraphics[width=\textwidth]{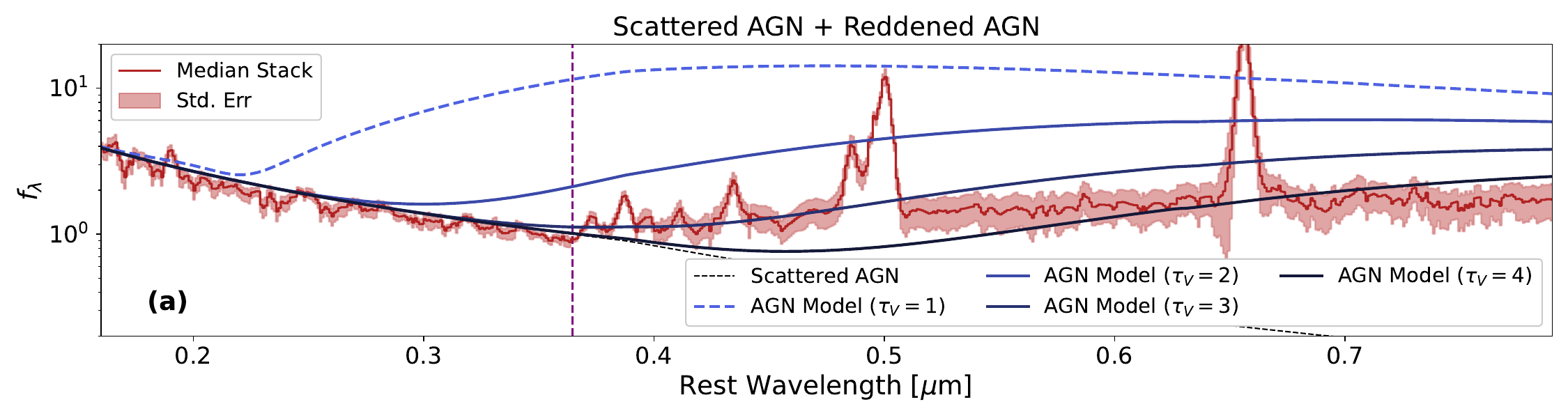}
    
    \includegraphics[width=\textwidth]{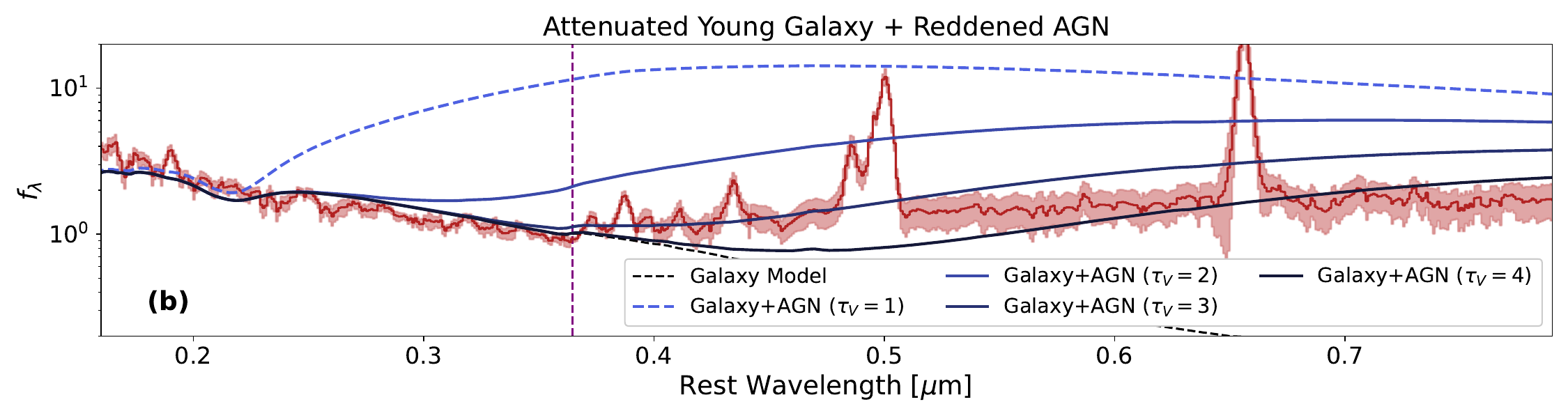}
    \includegraphics[width=\textwidth]{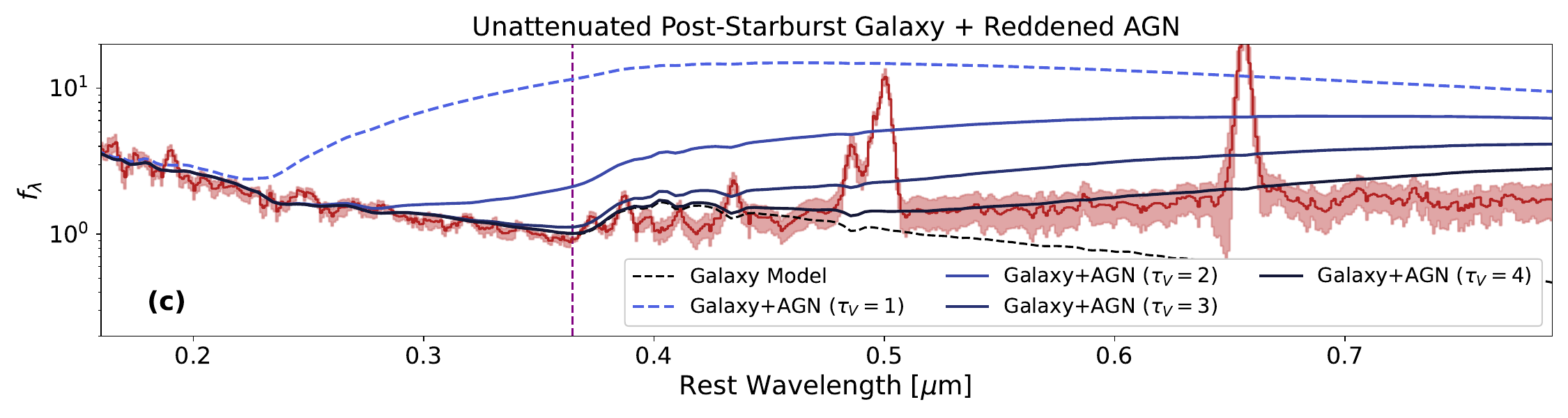}
    \includegraphics[width=\textwidth]{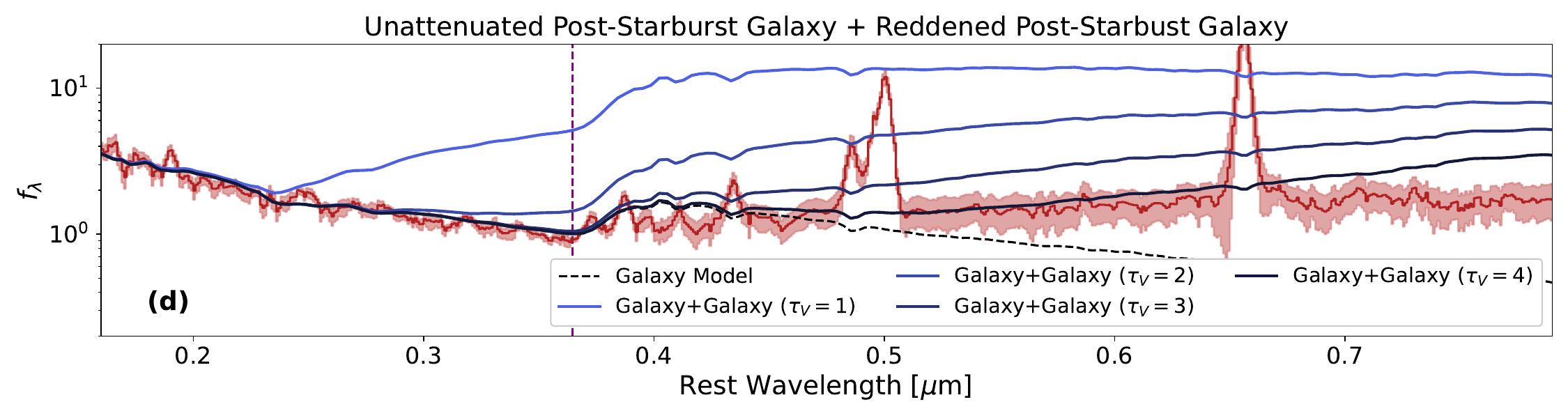}

    \caption{A median stack of the spectra (red) of all spectra with high-confidence slope changes. On the top, we show four AGN-AGN models, which are generated by scattering 1\% of the light from an intrinsic AGN continuum model \citep[][blue dashed line]{Temple2021} and attenuating the remainder of the light by the indicated $\tau_V$ using a \cite{Kriek2013} dust law with a dust index of $-1$. The AGN models are normalized so that the contribution from the scattered component is 1 at 3645 \angstromcomma, which is also where the spectra in the stack are normalized. In the next two rows, we show the same set of AGN models, but replacing the scattered light component with an $\tau_v\sim0.75$ star forming and dust-free post-starburst galaxies normalized to the UV. Models that meet the B-R$>0.95$ cut are shown as solid lines, and models that do not are shown as dashed. Finally, we show a two component galaxy model with the same unreddened post-starburst as the previous row, with the reddened AGN is replaced by the same stellar population undergoing varying reddening. In all cases, models \textit{can} produce change in slope at $3645$ \angstromcomma, but only with a very specific configuration of galaxy age, galaxy/AGN flux balance and reddening. It is unlikely that any two-component model where the two components originate from different physical origins would result in this uniform inflection location.}
    \label{fig:model_test}

\end{figure*}

\subsection{An inflection arising from the transition between a reddened AGN continuum and stellar continuum} \label{subsec:galaxy-agn}

Because the model of a reddened AGN continuum with a scattered light component fails to meet the criteria necessary to describe observed Little Red Dot spectra, we now turn to another commonly invoked model where the Little Red Dot spectral shape is caused by the intersection between a UV component that is dominated by galaxy light and a reddened AGN. This model has been shown to successfully replicate the continuum shape of spectroscopically confirmed Little Red Dots with broad lines \citep{Wang2024_BRD, Wang2024_UB, Ma2024_LRD}, though the exact range of contribution of galaxy and AGN light at a given wavelength--as well as the age, reddening, and mass of the underlying stellar component--can be highly uncertain.

We test two versions of this model. In the first, the underlying stellar population is young and moderately dusty (simulated as a 2 Myr old simple stellar population embedded in $\tau_V=0.25$). We treat the reddened quasar models identically to in Section \ref{subsec:agn-only}. In Figure \ref{fig:model_test}b, we show this model for the same range of AGN attenuation ($\tau_V=1-4$) as compared to our $2\sigma$ stack. As expected, this model fails to capture the true shape of Little Red Dots almost identically to the model where the UV results from an unobscured AGN; because there is no intrinsic inflection point in either of the intersecting models. The change in slope can occur at any wavelength, and the resultant breaks are not nearly as sharp as is observed in the stack.

It has therefore been suggested instead that the underlying stellar population that is producing the UV emission must be evolved enough that it lacks O and B stars, resulting in a Balmer break \citep{Wang2024_BRD, Wang2024_UB, Ma2024_LRD}, with the reddened AGN component picking up just redward of the Balmer break to produce the observed red continuum. Thus, we test a second implementation of the galaxy+AGN model by replacing the previous young simple stellar population instead with a dust-free 100 Myr old simple stellar population with light dominated by A stars. In Figure \ref{fig:model_test}c, we show this model compared to our stack, again with the same range of reddened quasar models. In contrast with the previous two models, this model is fully capable of replicating the UV continuum, break, and red continuum for the proper configuration of AGN attenuation and luminosity relative to the underlying stellar population. This model does suffer from the specificity required in this balance between AGN and galaxy light, but that may be partly due to selection--sources where the reddened AGN is totally dominant over the host galaxy would not have any significant UV component and would not be fit with any breaks in our model. It is therefore feasible, at least qualitatively, that ``V-shaped" Little Red Dots could result from the intersection of an evolved stellar population and a reddened AGN.

\subsection{An inflection resulting from a dusty evolved stellar population with some UV escape} \label{subsec:galaxy-galaxy}

It has also been suggested that the entire continuum in Little Red Dots can be ascribed to starlight, with standard SED modeling approaches adequately capturing the UV+optical with a young unattenuated population and the red end captured by dust-obscured stars \citep[e.g.,][]{Labbe2023UB,Akins2023,Williams2023, Akins2024, Wang2024_UB}. These models solve a number of problems present in the Galaxy+AGN model, including the lack of X-ray detections \citep{Furtak2024,Maiolino2024,Yue2024, Akins2024, Ananna2024} and flat mid-IR SEDs \citep[e.g.,][]{Williams2023, Wang2024_BRD, Akins2024}. It has also been suggested that the width of the broad Balmer emission in these models is fully consistent with the compact sizes of these sources and does not need to originate from an accretion disk \citep{Baggen2024}, lending credence to the idea that these systems could simply be the early-forming cores of present day elliptical galaxies \citep{Baggen2023}. While the source of ionizing radiation \citep[e.g.,][]{Ma2024_LRD} and the dust geometry \citep[Little Red Dot broad lines are consistent with significant attenuation, but the narrow lines are not, see][]{Brooks2024} for the Balmer emission presents an issue for such models, they merit exploration due to the natural connection between a break at H$_\infty$ and evolved stellar populations. 

In Figure \ref{fig:model_test}d, we demonstrate that this model can, similarly to the galaxy+AGN model, produce shapes that are broadly consistent with our Little Red Dot stack without necessitating significant tuning, even under the simplified assumption that the galaxy is a simple stellar population, where the UV originates from 1\% of stars that are unattenuated and the red continuum now comes from the remaining 99\% of galaxy light that sees a dust screen. Thus, it is indeed possible that the typical Little Red Dot spectral shape results from starlight alone, provided there is enough unattenuated light to produce the observed UV+break and that the reddened component is sufficiently dominant past the Balmer limit to ensure that the rest-frame B-R criteria is satisfied.

\subsection{An inflection resulting from a non-stellar $T=10^4$K hydrogen absorption process}

In the previous two sections, we demonstrated that the inflection at H$_\infty$ observed in Little Red Dots can be replicated by models that have sufficiently evolved stellar populations such that they intrinsically have Balmer breaks. However, A type stars are not the only place where Balmer absorption can occur, and in principle, a reservoir of $\sim10^4$ K gas associated with the accretion disk could produce a feature that looks remarkably similar to a stellar Balmer break by absorbing photons from the AGN continuum. One such model was proposed by \cite{Inayoshi2024}, wherein the accretion disk is embedded in extremely dense gas at this temperature. Such a model can account for the kinematically-offset absorption seen in the broad H$\alpha$ in some systems \citep[e.g.,][]{Wang2024_BRD,Matthee2024,Juodzbalis2024}. An intrinsic inflection in the accretion disk spectrum could also result from an opacity gap from hydrogen at the point where the disk temperature reaches $\sim10^4$ K \citep{Thompson2005}. While we do not explore these classes of models in detail, they (and other AGN models that can result in intrinsic breaks at H$_\infty$) merit further exploration given the clear empirical link to a preferred inflection location in Little Red Dots and the wide range in observed changes in spectral shape that occur at this very specific wavelength.

\section{Discussion and Conclusions} \label{sec:conclusions}

In this work, we leverage the red selection of RUBIES \citep{deGraaff2024_RUBIES}, supplementing with all public spectra hosted in the Dawn JWST Archive, to study the continuum shapes of extremely red sources with strong H$\alpha$ emission. We find that all unresolved sources that exhibit strong changes in spectral slope in the rest-optical \citep[which we consider to be ``V-shaped" Little Red Dots, e.g., ][]{Labbe2023LRD,Greene2024, Kocevski2024} consistently exhibit that inflection point very near to H$_\infty$, indicating an empirical connection between these changes in slope and $n=2$ hydrogen. The preference for an inflection point at this very specific wavelength rules out AGN-only models that account for the change with the intersection of a scattered light component and a highly reddened AGN component, similar to what is seen in more typical extremely red quasars that show no such preference for a specific break location \citep[e.g.,][]{Assef2016, Hamann2017}. This also rules out ``V-shaped" Little Red Dots resulting from the intersection between a young stellar component and an AGN, as these models would suffer from the same lack of a specific preferred inflection location as the aforementioned model.

In contrast, we demonstrate that models where an evolved stellar population dominates the UV through the rest optical and the reddened AGN component picks up just past the break \textit{can} produce this ``V-shape" at H$_\infty$, provided that the AGN is sufficiently reddened. For this scenario to be ubiquitous in the brightest Little Red Dots, there would need to be a link between the age of the stellar population (which must be $\gtrsim$50 Myr) and this particular reddened AGN SED. This model would also have clear observational consequences if the duty cycle of Little Red Dots is $<$100\%, because the AGN turning off (and therefore, Little Red Dots losing the reddened AGN continuum that accounts for the rising SED redward of the break) would leave behind a post-starburst core that would remain detectable as compact, evolved stellar population, unless it simultaneous restarts forming stars coincidentally with the AGN shutting down. The massive quiescent galaxy identified in \cite{Weibel2024} at z=7.3 \textit{does} strongly match the spectral shape of the \cite{Furtak2024} Little Red Dot (at very similar redshift) up to $\sim4000 \ \AA$, making such a connection plausible. However, it remains to be seen whether the number densities work out for this model to hold for the full population of high-z Little Red Dots.

Additionally, the AGN in composite models is difficult to square with the deficiency of hard X-ray photons from a hot corona \citep{Furtak2024, Maiolino2024, Yue2024, Akins2024, Ananna2024} and the lack of a NIR rise associated with a dusty torus \citep{Williams2023, Akins2024, Wang2024_BRD}.
Both these problems could potentially be resolved if the AGN torus is cold \citep[e.g.,][]{Li2024_LRD}, possibly due to a super-Eddington accretion disk that is redder than typical quasar spectra \citep[e.g.,][]{Volonteri2017,Lambrides2024}. Indeed X-ray weakness has also been observed in extremely red quasars, though they have more typical hot dust tori that are detected in the mid-IR \citep{Hamann2017, Ma2024_ERQ}. 
However, a precise configuration of stellar age, galaxy and AGN dust, and relative luminosity is required for such models to achieve the observed breaks without violating some constraint from the rest-optical through the sub-mm.

The class of models that produce these breaks through stellar emission alone also present a number of unresolved problems. Because they require significant dust (and often dust-obscured star formation) to properly describe the red continuum, they also predict significant sub-mm emission from cold dust \citep[e.g.,][]{Labbe2023LRD}, and to date no Little Red Dot has been detected in the FIR, even in stacks \citep[e.g.,][]{Labbe2023LRD,Williams2023,Akins2024}, implying very high dust temperatures and low dust masses \citep{Casey2024}. There is still some room for a red continuum that is dominated by reddened older stellar populations, which could alleviate tension with FIR non-detections, but these models produce a factor of $\sim$100 too few ionizing photons to account for the observed Balmer emission, even if the kinematics and continuum shape are consistent with starlight \citep{Ma2024_LRD}.  

The stellar masses implied by these galaxy-only models for the brightest Little Red Dots also present a clear problem for galaxy evolution theory, with number densities that imply baryon conversion efficiencies of $\sim100\%$ \citep[e.g.,][]{BoylanKolchin2023}--though we note that such early-forming solutions appear at present to describe the star formation histories of some of the most extremely massive quenched galaxies at $z=3-5$ \citep{Glazebrook2024, Carnall2024, deGraaff2024_QG}. Furthermore, Little Red Dots are unresolved in the rest-optical even when subjected to significant lensing \citep[with sizes $<30$ pc, see][]{Furtak2024}.  The combination of these compact sizes with the aforementioned high stellar masses results in extreme stellar densities that essentially require that they be the progenitors of the dense cores of massive elliptical galaxies. Given that the most massive quiescent galaxies observed at $z=4-7$ \textit{do} have extended stellar mass profiles \citep[with $r_e\sim$ a few hundred pc, see][]{Carnall2023b, deGraaff2024_QG, Weibel2024}, it also implies that if this evolutionary link holds that Little Red Dots must rapidly grow a stellar envelope around their core to match observations, potentially while losing mass in their cores to avoid being overdense \citep{Baggen2024}. Some of this tension with these uncomfortably large masses and densities can potentially be reduced while remaining consistent with local elliptical cores with modifications to the initial mass function \citep{VanDokkum2024}, but it is essential that better constraints be placed on the number densities of both Little Red Dots and early quiescent populations to test whether a progenitor-descendant evolutionary pathway between the two is plausible.

Ultimately, we remain agnostic as to which specific Balmer break process is occurring in Little Red Dots. While we demonstrate that stellar populations that are dominated by A-type stars can describe the rest-UV-to-optical spectral shape of Little Red Dots, it also remains possible that some intrinsic AGN-only process involving $\sim 10^4$ K hydrogen absorption could produce the observed shapes. Whatever physical process drives the transition in spectral slope, we stress that the association with this specific inflection point of H$_\infty$ is empirically a necessary component of describing the rest-optical continuum shape of Little Red Dots. Future work leveraging this information along with the full suite of ancillary observations from the X-ray to the FIR/radio will be essential to reconcile the high number densities and unique SEDs of these new and exciting high-z sources.

\facilities{JWST (NIRCam, NIRSpec)}

\software{Astropy \citep{astropy2013, astropy2018, astropy2022}, \texttt{STScI} JWST Calibration Pipeline \citep[\url{https://jwst-pipeline.readthedocs.io/}][]{Rigby2023_jwstpipeline}, \texttt{grizli} \citep[\url{github.com/gbrammer/grizli}]{Brammer2023_grizli}, \texttt{msaexp} \citep[\url{https://github.com/gbrammer/msaexp}][]{Brammer2023_msaexp},
Matplotlib \citep{Hunter:2007}, Flexible Stellar Population Synthesis \citep{Conroy2009, Conroy2010}, SEDPy \cite{sedpy2019}}

\acknowledgements

 This work is based in part on observations made with the NASA/ESA/CSA \textit{James Webb Space Telescope}. The data were obtained from the Mikulski Archive for Space Telescopes at the Space Telescope Science Institute, which is operated by the Association of Universities for Research in Astronomy, Inc., under NASA contract NAS 5-03127 for JWST. The JWST data presented in this article were obtained from the Mikulski Archive for Space Telescopes (MAST) at the Space Telescope Science Institute. All of the data products presented herein were retrieved from the Dawn JWST Archive (DJA). DJA is an initiative of the Cosmic Dawn Center, which is funded by the Danish National Research Foundation under grant No. 140. We express gratitude toward the members of the GTO, GO, and DDT teams whose public data we utilized in this work.
 
 Support for this work was provided by The Brinson Foundation through a Brinson Prize Fellowship grant. Support for program \#4233 was provided by NASA through a grant from the Space Telescope Science Institute, which is operated by the Association of Universities for Research in Astronomy, Inc., under NASA contract NAS 5-03127. This research was supported by the International Space Science Institute (ISSI) in Bern, through ISSI International Team project \#562. DS acknowledges helpful conversations with Xiaohui Fan and Jared Siegel that contributed to the quality of this work, in addition to aesthetic sign-off from Stephanie Permut on the colors of figures. TBM was supported by a CIERA fellowship.
The work of CCW is supported by NOIRLab, which is managed by the Association of Universities for Research in Astronomy (AURA) under a cooperative agreement with the National Science Foundation.  

\bibliography{LRD_Breaks}

\end{document}